# Microwave photonic frequency measurement and time-frequency analysis: Unlocking bandwidths over hundreds of GHz with a 10-nanosecond temporal resolution

Taixia Shi,[a,b] Chi Jiang,[a,c] Chulun Lin,[a,c] Fangyi Yang,[a,c] Yiqing Liu,[a,c] Fangzheng Zhang,[b] and Yang Chen[a,c,*]

[a] Shanghai Key Laboratory of Multidimensional Information Processing, School of Communication and Electronic Engineering, East China Normal University, Shanghai 200241, China
[b] National Key Laboratory of Microwave Photonics, Nanjing University of Aeronautics and Astronautics, Nanjing 210016, China
[c] Engineering Center of SHMEC for Space Information and GNSS, School of Communication and Electronic Engineering, East China Normal University, Shanghai 200241, China
*Correspondence to: ychen@ce.ecnu.edu.cn

**ABSTRACT**
Fast and broadband spectrum sensing is an essential component in cognitive radio systems, intelligent transportation systems, electronic warfare systems, etc. However, traditional electronic-based solutions have a trade-off among the analysis bandwidth, temporal resolution, and real-time performance. In comparison, microwave photonic solutions can overcome the trade-off at the cost of frequency accuracy and resolution. Nevertheless, the reported microwave photonic solutions suffer from a very poor frequency resolution and impose extremely high requirements on hardware when the analysis bandwidth is close to or greater than 100 GHz. Here, we show a microwave photonic frequency measurement and time-frequency analysis method, which is implemented by dispersion-based frequency-to-time mapping and assisted by a specially designed V-shape linearly frequency-modulated signal and a duty-cycle-enabling technique. Compared with the reported microwave photonic solutions, the hardware requirements are greatly reduced when achieving similar performance conditions. Using a total dispersion of −6817 ps/nm and a V-shape linearly frequency-modulated signal with a bandwidth of 31.6 GHz and a duty cycle of 1/4, we achieve an ambiguity-free analysis bandwidth of 252.8 GHz, a corresponding temporal resolution of 13.75 ns and a frequency resolution of 1.1 GHz. The temporal resolution can be improved to 6.875 ns when the duty cycle is changed to 1/2, while the analysis bandwidth in this case is 126.4 GHz.

## 1. Introduction

Fast, accurate, and broadband microwave frequency measurement and time-frequency analysis can be widely applied in cognitive radio systems [1–3], intelligent transportation systems [3], and electronic warfare systems [4, 5]. Microwave frequency measurement and time-frequency analysis implemented by traditional electronic-based methods have very good measurement accuracy and resolution. However, when measuring broadband signals, severe limitations in measurement real-time performance and reconfigurability are often encountered, which is attributed to the inherent electronic bottleneck.

Microwave photonics [6, 7] utilizes photonic technology to generate, process, and transmit microwave signals. In terms of microwave frequency measurement and time-frequency analysis, it offers significantly better real-time performance and reconfigurability compared to traditional electronic-based solutions, particularly when measuring high-frequency and large bandwidth signals [8, 9]. Therefore, it is recommended to combine the microwave photonic approach with the electronic-based approach: The microwave photonic approach is utilized for rapid and "coarse" localization of the signal under test (SUT) across an ultra-wide frequency band, which is then handed over to the electronic-based approach for precise analysis of the SUT within the specific frequency bands where SUTs are present [9, 10]. By combining the two approaches, it is expected to achieve real-time and high-accuracy analysis of the ultra-wide spectrum.

Microwave frequency measurement and time-frequency analysis can be implemented by frequency-to-power mapping [11–13], frequency-to-space mapping [10, 13], and frequency-to-time mapping (FTTM). The FTTM-based method has gained widespread attention in recent years, primarily due to its ease of signal measurement in complex and non-single-tone spectral environments. Microwave frequency measurement [14–21] and time-frequency analysis [10, 22–30] via FTTM can be accomplished mainly in two ways, including FTTM via optical frequency-sweeping and filtering [10, 14–17, 22, 23] and FTTM via dispersion [18–21, 24–30]. The former based on optical frequency-sweeping and filtering employs a high-speed frequency-sweep signal to convert the SUT into a series of chirped optical signals. These chirped optical signals are then filtered by a fixed narrowband optical filter. As a result, the chirped optical signals corresponding to different frequencies of the SUT will sweep through the filter at different times, generating pulses that appear at distinct moments. By measuring the appearance times of these pulses, the frequency of the SUT can be determined. In addition to the implementation with a chirped optical signal and fixed optical filter mentioned above [10, 14–16, 22, 23], the former can also be achieved by a fixed optical signal combined with a frequency-sweep optical filter, and the principle is similar [17]. The primary limitation of this kind of method lies in its temporal resolution, which is typically only on the order of microseconds, making it difficult to meet the requirements of applications with higher real-time requirements. Besides, the analysis bandwidth is commonly limited to tens of GHz. Although a method based on channelization [10] has been proposed, theoretically allowing for a significant increase in temporal resolution and analysis bandwidth through the number of channels, this comes at the cost of significantly increased system complexity.

The latter based on dispersion mostly works in the following manner: A periodic temporal quadratic phase is loaded on the optical signal that is used to carry the SUT; by matching the temporal quadratic phase and the dispersion value of the dispersion medium (DM), the waveform carrying the SUT which is dispersed over a large time scale is compressed into narrow pulses in the temporal domain. In [20] and [21], the broadband optical signal with a temporal quadratic phase is obtained by stretching the pulses from a mode-locked laser (MLL) using a DM with a large dispersion value. To simplify the structure used in [20] and [21], the DM for stretching pulses is removed based on the Talbot design [26]. Furthermore, both the MLL and the DM for stretching pulses can be replaced by a continuous-wave (CW) laser diode (LD), a phase modulator (PM), and an auxiliary signal generated from an arbitrary waveform generator (AWG), based on optical time-lens [25, 28, 29] or fractional Talbot designs [30]. In [29], a very large maximum ambiguity-free analysis bandwidth of 448 GHz, as well as an analysis bandwidth of 5 THz with possible spectral ambiguities, can be achieved. Although an analysis bandwidth of 5 THz is achieved in [29] by using an unwrapping technique, in practical applications, SUT rarely possesses solely monotonic time-frequency characteristics, making it difficult to use for the analysis of complex real-world signals. For the ambiguity-free 448-GHz analysis bandwidth achieved in [29], the frequency resolution is very poor at only 16 GHz and a very high requirement on hardware must be satisfied, including a 30-GHz bandwidth PM with a low half-wave voltage $V_\pi$ of 2.5 V, a 120-GSa/s sampling rate AWG, a time-lens signal with an amplitude of $7V_\pi$, and a 500-GHz bandwidth optical sampling oscilloscope (OSC). It is crucial to note that the system's reliance on a 500-GHz bandwidth optical sampling OSC inherently restricts its applicability to solely measuring periodic signals. However, in practical applications, the overwhelming majority of SUTs exhibit non-periodic characteristics. If a real-time OSC is used instead, the frequency resolution of the system in [29] will be further decreased on the basis of 16 GHz due to the inability to reach the 500-GHz bandwidth. To avoid the high hardware requirements in [29] and to gain the ability to measure arbitrary non-periodic signals, the fractional Talbot effect is employed with the assistance of a temporal Talbot array illuminator signal [30], achieving an analysis bandwidth of 92 GHz by using an AWG with a 92-GSa/s sampling rate, a temporal Talbot array illuminator signal with an amplitude of $2V_\pi$, and a 28-GHz OSC. Facing the demand for future ultra-wideband applications, there is an urgent need for a new method that can achieve an analysis bandwidth of over 100 GHz with lower hardware requirements, specifically lower auxiliary signal amplitude, lower sampling rate requirements for AWG, and lower bandwidth requirements for OSC.

In this work, a microwave photonic frequency measurement and time-frequency analysis method is proposed, which significantly reduces the aforementioned hardware requirements while unlocking an analysis bandwidth of over hundreds of GHz, with a temporal resolution of around 10 ns. The key to achieving the ultra-wide ambiguity-free analysis bandwidth lies in the use of a specially designed V-shape linearly frequency-modulated (LFM) signal combined with a duty-cycle-enabling technique: The specially designed V-shape LFM signal enables the system to achieve an analysis bandwidth that is twice that of the V-shape LFM signal itself; The duty-cycle-enabling technique provides the system with the potential for further significant expansion of the analysis bandwidth to even THz. An experiment is performed. An ambiguity-free analysis bandwidth of 252.8 GHz is achieved by employing a duty cycle of 1/4 on the V-shape LFM signal, the corresponding temporal resolution is 13.75 ns, and the frequency resolution is 1.1 GHz by using a 64-GSa/s AWG, a 20-GHz OSC, and a total dispersion of −6817 ps/nm. The temporal resolution can be improved to 6.875 ns when the duty cycle is changed to 1/2, while the analysis bandwidth in this case is 126.4 GHz.

## 2. Principle and experimental setup

### 2.1 Experimental setup

The schematic diagram of the proposed microwave photonic frequency measurement and time-frequency analysis method is shown in Fig. 1. A CW light wave with a power of 6 dBm and a wavelength of 1546.6 nm is emitted by an LD (ID Photonics CoBriteDX1-1-

C-H01-FA) and then sent to a dual-parallel Mach–Zehnder modulator (DP-MZM, Fujitsu FTM7961EX) via an optical isolator and a polarization controller. The DP-MZM is biased as a carrier-suppressed single-sideband (CS-SSB) modulator [10]. Two specially designed V-shape LFM signals are generated using two different channels of an AWG (Keysight M8195A, 64 GSa/s). The two V-shape LFM signals have a ±90° phase difference as shown in Figs. 1(a) and (b) and possess a bandwidth $f_b$ of 31.6 GHz, a period $T$ of 3.4375 ns, and a negative chirp rate $k$ of −18.385 GHz/ns. They are then sent to the RF ports of the DP-MZM after being amplified by a dual-channel electrical amplifier (EA1, Centellax OA4SMM4). Due to the opposing 90° phase difference between the two V-shaped LFM signals during the intervals of 0–$T$/2 and $T$/2–$T$ within one period $T$, when the duty-cycle-enabling technique is not applied as shown in the left figures of Figs. 1(a) and (b), a periodic chirped optical signal with a bandwidth of 2$f_b$ is generated, as shown in the left figure of Fig. 1(c). When the duty-cycle-enabling technique is introduced, for example, the duty cycle is 1/2 as shown in the right figures of Figs. 1(a) and (b), a duty cycle of 1/2 is also introduced to the generated chirped optical signal as shown in the right figure of Fig. 1(c).

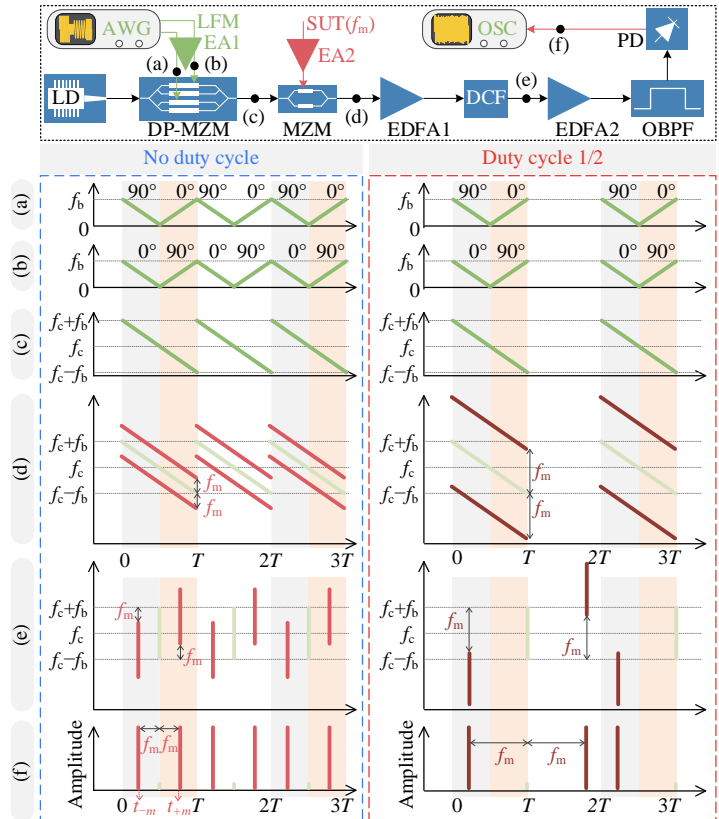

Fig. 1. Schematic diagram of the proposed microwave photonic frequency measurement and time-frequency analysis method. (a)–(f) Schematic of the signals at different locations in the system diagram. LD, laser diode; DP-MZM, dual-parallel Mach–Zehnder modulator; AWG, arbitrary waveform generator; LFM, linearly frequency-modulated; EA, electrical amplifier; MZM, Mach–Zehnder modulator; SUT, signal under test, EDFA, erbium-doped fiber amplifier; DCF, dispersion compensating fiber; OBPF, optical bandpass filter; PD, photodetector; OSC, oscilloscope.

The optical signal from the DP-MZM is sent to a Mach–Zehnder modulator (MZM, Fujitsu FTM7938EZ) which is biased at the minimum transmission point for carrier-suppressed double-sideband (CS-DSB) modulation. The SUT is applied to the RF port of the MZM via EA2 (Centellax OA4MVM3), where it is converted to a chirped optical signal as shown in Fig. 1(d) when considering a single-frequency SUT input. The optical signal from the MZM is amplified by an erbium-doped fiber amplifier (EDFA1, Amonics AEDFA-PA-35-B-FA) and then sent to a dispersion compensating fiber (DCF) with a total dispersion of around −6817 ps/nm at the wavelength of 1546.6 nm. Since the chirp rate of the V-shape LFM signal and the dispersion value are precisely matched, the chirped signal carrying the SUT will be compressed to optical pulses, as shown in Fig. 1(e). To determine the appearance time of the optical pulses, the optical pulses from the DCF are amplified by EDFA2 (Max-Ray EDFA-PA-35-B), filtered by an optical bandpass filter (OBPF, WL Photonics WLTF-BA-U-1550-100) with a bandwidth of around 350 GHz and a center wavelength around 1546.6 nm, and then injected into a photodetector (PD, u2t MPRV1331A) to convert them to electrical pulses, as shown in Fig. 1(f). The electrical pulses from the PD are monitored by an oscilloscope (OSC, LeCroy WaveMaster 820Zi-B, 80 GSa/s) with a maximum 3-dB bandwidth of 20 GHz. For any frequency of the SUT, the two chirped optical sidebands generated as shown in Fig. 1(d) will be converted into a pair of optical pulses. The frequency information of the SUT can then be obtained by measuring the time difference between the two pulses or measuring the appearance time of one pulse in one period. By accumulating the frequency measurement results over multiple periods, the time-frequency information of the SUT can also be obtained.

2.2 Principle of the FTTM process

Then, a single-tone SUT centered at $f_m$ is employed to analyze the principle of the FTTM process. Assuming the center frequency of the CW light wave is $f_c$, the start frequency and chirp rate of the +1st-order chirped optical sideband of the SUT are $f_c + f_b + f_m$ and $k = -2f_b/T$. Thus, the complex envelopes of the +1st-order chirped optical sideband of the SUT can be expressed as

$$E_1(t) \propto \exp\left(j2\pi f_s t + j\pi k t^2\right), \tag{1}$$

where $f_s = f_b + f_m$. After passing through the DCF, the optical signal can be expressed as

$$\begin{aligned} E_2(t) &\propto \exp\left(j2\pi f_s t + j\pi k t^2\right) * \exp\left(\frac{jt^2}{2\beta_2 L}\right) \\ &= \int_{-\infty}^{+\infty} \exp\left(j2\pi f_s \tau + j\pi k \tau^2\right) \exp\left[\frac{j(t^2 - 2t\tau + \tau^2)}{2\beta_2 L}\right] d\tau, \end{aligned} \tag{2}$$

where $\beta_2 = -\lambda^2 D/(2\pi c)$ is the dispersion coefficient, $L$ is the length of the DCF, $\lambda$ denotes the wavelength corresponding to $f_c$, $D$ represents the dispersion parameter, $c$ is the light speed in vacuum, and the symbol "*" means convolution. To compress the chirped optical sideband to a narrow pulse, the chirp rate of the chirped optical signal should be matched with the total dispersion ($DL$). The matching relationship can be expressed as

$$k = -\frac{1}{2\pi\beta_2 L}$$

$$= \frac{c}{\lambda^2 DL}.$$

(3)

Under this specific circumstance given by Eq. (3), when only one sweep period $T$ is considered, Eq. (2) can be further given by

$$\begin{aligned}
E_2(t) &\propto \int_0^T \exp(j2\pi f_s \tau) \exp\left(\frac{jt^2 - j2t\tau}{2\beta_2 L}\right) d\tau \\
&= \int_0^T \exp[j\tau(2\pi f_s + 2\pi kt)] \exp(-j\pi kt^2) d\tau \\
&= \exp(-j\pi kt^2) \frac{\exp[j2\pi(f_s + kt)T] - 1}{j2\pi(f_s + kt)} \\
&= \exp(-j\pi kt^2) \exp[j\pi(f_s + kt)T] Sa[\pi(f_s + kt)T] T,
\end{aligned}$$

(4)

where $Sa(\cdot)$ denotes the sampling function. As can be seen from Eq. (4), the +1st-order chirped optical sideband of the SUT is compressed into a temporal-domain pulse and takes on the shape of a sampling function. When $f_s + kt = 0$ is established, the peak of the compressed pulse is obtained through Eq. (4). When the frequency of the single-tone SUT gradually increases from 0 to $f_b$, $f_s$ also gradually increases accordingly, and the change in $f_s$ from when the SUT frequency is 0 to when it reaches $f_b$ is $f_b$. Therefore, the peak of the compressed pulse also gradually shifts when the SUT frequency increases from 0 to $f_b$, with a variation of $-f_b / k = T/2$. Thus, the SUT frequency can be expressed as

$$f_m = -k(t_{+m} - T/2),$$ 

(5)

where $t_{+m} \in [T/2, T]$ is a relative position of the pulse in a period $T$. Similarly, for −1st-order chirped optical sideband of the SUT, the SUT frequency can be expressed as

$$f_m = k(t_{-m} - T/2),$$

(6)

where $t_{-m} \in [0, T/2]$ is also a relative position of the pulse in a period $T$.

2.3 Analysis bandwidth, resolution, and duty cycle

When the duty-cycle-enabling technique is not employed, according to Fig. 1, Eqs. (5) and (6), for any frequency $f_m$, a pair of symmetric pulses can be respectively generated in time duration of 0–$T/2$ and $T/2$–$T$. The corresponding frequency analysis range is from $-f_b$ to $f_b$. Indeed, the symmetry of the analysis results and the frequency analysis range is mainly induced by the CS-DSB modulation implemented in the MZM in Fig. 1, which introduces two symmetric chirped optical sidebands of the SUT. If the MZM in Fig. 1 is replaced by another DP-MZM and CS-DSB modulation is replaced by CS-SSB modulation, the symmetry of the analysis results will be eliminated, allowing the system to truly have an analysis bandwidth of $2f_b$ (0 to $2f_b$). If the SUT frequency exceeds $f_b$ using CS-DSB modulation or $2f_b$ using CS-SSB modulation, after the chirped optical sidebands of the SUT are compressed in the DCF, crosstalk will emerge between optical pulses from different sweep periods, making it impossible to determine the signal frequency based on the pulse positions. Thus, the ambiguity-free analysis bandwidth of the proposed system can be considered to be $2f_b$ when the duty-cycle-enabling technique is not employed.

When the duty-cycle-enabling technique is employed, a duty cycle of $p$ is applied in the V-shape LFM signal. In this case, after the chirped optical sidebands of the SUT passes through the DCF, although when the SUT frequency is very high and the pulse corresponding to a sweep period may enter its adjacent periods, the introduction of the duty cycle creates some gaps in the temporal domain, which can avoid the issue of pulse crosstalk mentioned above. Therefore, the ambiguity-free analysis bandwidth can be extended to $2f_b/p$ ($-f_b/p$ to $f_b/p$ or 0 to $2f_b/p$). For instance, when the duty cycle of the V-shape LFM signal is 1/2, the ambiguity-free analysis bandwidth is $4f_b$, as shown in the right parts of Figs. 1(d)–(f).

The temporal resolution of the system is determined by how long the system analyzes the whole analysis bandwidth one time. Therefore, when the duty-cycle-enabling technique is not enabled, the temporal resolution $r_t$ is $T$. When the duty-cycle-enabling technique is enabled and the duty cycle is set to $p$, the temporal resolution deteriorates to $T/p$.

The frequency resolution of the system can be determined by the frequency width associated with the zero-crossing width of the pulse after FTTM. According to Eq. (4), the zero-crossing width of the pulse is $2/kT$. If the frequency resolution of the system is defined as the frequency width associated with the full-width at half maximum (FWHM) of the pulse after FTTM, according to Eq. (4), the FWHM of the pulse is around $|1.2068/kT|$. No matter how the frequency resolution is defined, a unified conclusion is that the frequency resolution of the system is closely related to the bandwidth and chirp rate of the chirped optical signal. Using the sweep chirp rate of the chirped optical signal and converting the pulse width into frequency width, the frequency resolution of the system, when the FWHM of the pulse is used to determine the frequency resolution, can be expressed as

$$r_f \approx 1.2068/T. \tag{7}$$

It is important to note that Eq. (7) is the theoretical frequency resolution, and the prerequisite for achieving this resolution in a real-world system is that the PD used for optical-to-electrical conversion and the OSC used for signal acquisition have sufficient bandwidth to acquire the optical pulse given in Eq. (4) without distortion. Besides, regardless of whether the duty-cycle-enabling technique is used or not, the frequency resolution is always defined by Eq. (7), which is different from the case of temporal resolution. It is noteworthy that the trade-off between temporal resolution and frequency resolution, as dictated by Heisenberg's uncertainty principle, can also be deduced from Eq. (7). Specifically, the higher the temporal resolution, the lower the frequency resolution, and conversely, the better the frequency resolution, the poorer the temporal resolution becomes.

## 3. Experimental results

3.1 Measurement without duty-cycle-enabling technique

First, the duty-cycle-enabling technique is disabled in the system. The analysis bandwidth in this case is $2f_b = 63.2$ GHz and the sweep period $T$ is 3.4375 ns. The SUT is generated by a third channel of the AWG with an output power of −16 dBm. Before implementing the time-frequency analysis, the noise floor of the OSC, as well as the noise introduced by the PD, is studied when the sampling rate and 3-dB bandwidth of the OSC is 80 GSa/s and 20 GHz, with the noise waveform shown in Figs. 2(a) and (b). The root mean square value of the noise when the PD is disconnected from the OSC is 0.85 mV. When the PD is connected to the OSC, the root mean square value of the noise is 1.50 mV. Figure 2(c) shows the time-frequency diagram obtained from the raw data captured by the OSC when the SUT is an LFM signal sweeping from 1 GHz to 30 GHz. Due to the influence of output noise from the OSC and PD, the background noise in the time-frequency diagram is relatively high. The raw data is processed using the clipping method, adjusting all waveforms below 0 V to 0 V, and the time-frequency diagram after processing is shown in Fig. 2(d). It can be seen that the noise can be well suppressed without affecting the SUT time-frequency analysis results. In the following demonstrations, the raw data is all processed in this way to obtain a better time-frequency diagram. The waveforms used to draw the time-frequency diagrams in Figs. 2(c) and (d) are respectively given in Fig. 2(e). There is no pulse seen in Fig. 2(e) because it contains 1024 sweep periods. Additionally, the amplitude of the pulses, i.e., the envelope of all the pulses shown in Fig. 2(e), can reflect the variation of the power spectrum of the SUT with frequency. Figure 2(f) shows the electrical spectrum of an LFM signal generated from the AWG from 0 to 32 GHz. The spectrum is measured by using an electrical spectrum analyzer (Ceyear 4052E). It can be seen that the envelope of the spectrum in Fig. 2(f) is highly consistent with the envelope of the waveform in Fig. 2(e). Further combined with Fig. 2(d), it can be seen that the brightness of the time-frequency diagram also matches well with the power variation with frequency in the spectrum in Fig. 2(f).

Since we applied CS-DSB modulation to the SUT, both sidebands of SUT are analyzed, resulting in the time-frequency diagrams shown in Figs. 2(c) and (d). This leads to an analysis bandwidth from −31.6 GHz to 0 and from 0 to 31.6 GHz, totaling 63.2 GHz. However, due to CS-DSB modulation, the information in these two spans is redundant. This redundancy can be avoided by using CS-SSB modulation, enabling a true analysis bandwidth of $2f_b = 63.2$ GHz. Besides, the amplitude of the waveforms is normalized and then used for drawing all the time-frequency diagrams in this work, while it is not normalized when only the waveform is given.

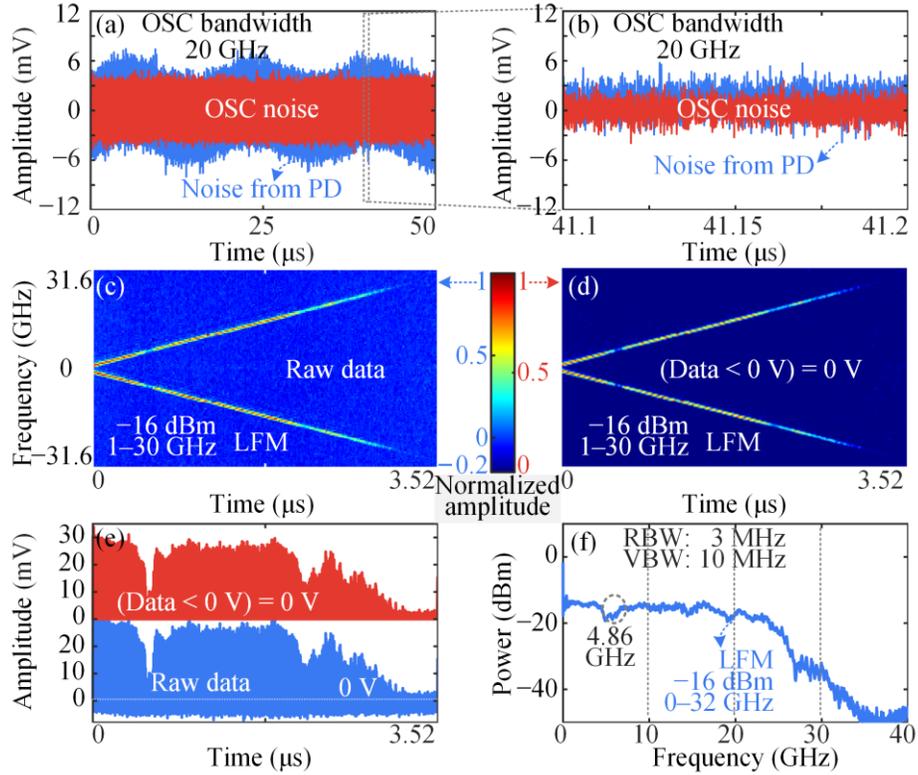

Fig. 2. (a) Waveforms of the noise captured by the OSC with and without PD connection. (b) Zoomed-in view of the black-dotted rectangular area in (a). Time-frequency diagrams of an LFM signal obtained by the system (c) without and (d) with noise exclusion. (e) Waveforms corresponding to (c) and (d). (f) Electrical spectrum of the LFM signal from the AWG after amplification by EA1.

To further demonstrate the time-frequency analysis capability of the proposed system for more complex signals, a nonlinearly frequency-modulated (NLFM) signal with a frequency range of 1–30 GHz, a V-shape LFM signal with a frequency range of 1–30 GHz, a signal with a sinusoidal time-frequency diagram with a frequency range of 1–30 GHz, and a signal with "ECNU" time-frequency diagram with a frequency range of 3–23 GHz are used as the SUTs within a time length of 3.52 μs. Here, "ECNU" is the abbreviation for "East China Normal University". The corresponding time-frequency analysis results are shown in Figs. 3(a)–(d). It can be seen that the time-frequency characteristics of different SUTs can be well recovered in an analysis bandwidth of 63.2 GHz.

To evaluate the analysis accuracy of the proposed system, microwave frequency measurement is studied. Five single-tone SUTs with frequencies of 5 GHz, 10 GHz, 15 GHz, 20 GHz, and 25 GHz are measured 2000 times via 2000 sweep periods. The 2000 measurement results are shown in Fig. 3(e). It can be observed that for any single-tone signal, the measurement result is always one of two adjacent frequency values, which correspond to two adjacent sampling points of the generated pulse. The difference between these two frequency values corresponds to the sampling interval multiplied by the chirp rate of the chirped optical signal (V-shape LFM signal). Figure 3(f) displays five different waveforms obtained from individual measurements of five single-tone signals. The average measurement values of the five frequencies obtained from 2000 measurements are 4.9324 GHz, 9.9955 GHz, 15.0860 GHz, 20.1121 GHz, and 25.0781 GHz, respectively, while the corresponding standard deviations are 44.0 MHz, 19.2 MHz, 80.5 MHz, 26.0 MHz, and 111.5 MHz, respectively. The measurement errors and the corresponding error bars are shown in Fig. 3(g). Based on the above experiments, The frequency measurement and time-frequency analysis capabilities of the proposed method have been validated.

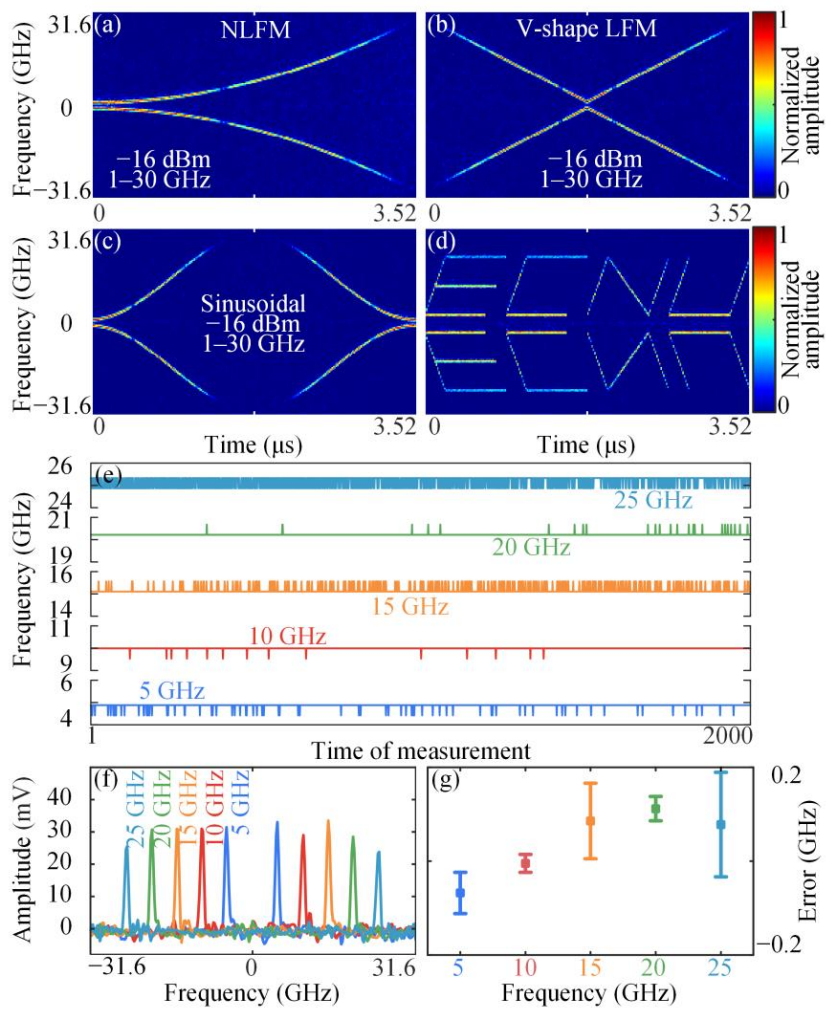

Fig. 3. Measured time-frequency diagrams of (a) an NLFM signal, (b) a V-shape LFM signal, (c) a signal with a sinusoidal time-frequency diagram, and (d) a signal with an "ECNU" time-frequency diagram. (e) Measurement results of 2000 frequency measurements on five single-tone signals. (f) Waveforms of a single frequency measurement. (g) Measurement errors of the five single-tone signals.

3.2 Extending the analysis bandwidth via duty-cycle-enabling technique

Then, the duty-cycle-enabling technique is introduced to further extend the ambiguity-free analysis bandwidth of the proposed system. In this study, the duty cycle of the V-shape LFM signal applied to the DP-MZM is changed to 1/2, 1/3, and 1/4, the corresponding temporal resolutions are 6.875 ns, 10.3125 ns, and 13.75 ns, and the corresponding time-frequency analysis results are shown in Figs. 4 (a)–(d), (e)–(h), and (i)–(l), respectively. It can be seen that the analysis bandwidth is extended to 126.4 GHz, 189.6 GHz, and 252.8 GHz when the duty cycles are 1/2, 1/3, and 1/4. Theoretically, the analysis bandwidth can be further increased by further reducing the duty cycle. Since we do not have a signal source with such high frequency, in this experiment, we generate the ±3rd-order and even

±5th-order harmonics during the optical modulation process by increasing the power of the SUT to equivalently verify the above-mentioned ultra-wide analysis bandwidth. As shown in Fig. 4(l), when a single-tone signal at 15 GHz is used, the ±5th-order harmonics are successfully marked in the obtained time-frequency diagram, indicating that a 150-GHz analysis bandwidth is confirmed. Actually, the analysis bandwidth is 252.8 GHz, as shown in Figs. 4(i)–(l), however, we cannot generate such high-frequency harmonics. Figure 4(m) displays the temporal waveforms corresponding to the head, middle, and tail portions of the time-frequency diagram in Fig. 4(i). The corresponding frequency information on the time-frequency diagram is reflected as pulses at different positions on the temporal-domain waveform.

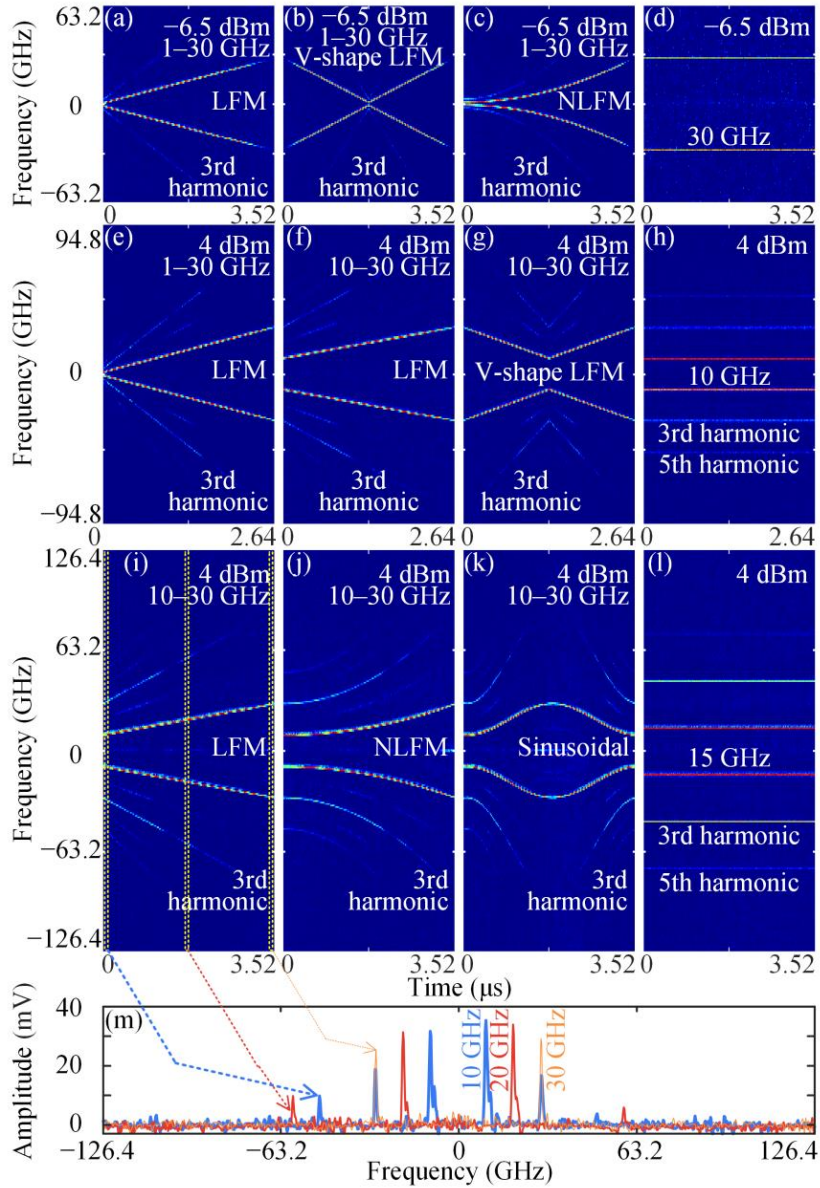

Fig. 4. Time-frequency analysis results when the duty cycle of the V-shape LFM signal applied to the DP-MZM is set to (a)–(d) 1/2, (e)–(h) 1/3, and (i)–(l) 1/4. (m) Temporal waveforms corresponding to the head, middle, and tail portions of the time-frequency diagram in (i).

3.3 Frequency resolution

The frequency resolution of the proposed system is then investigated. First, a two-tone test is carried out. As indicated in Eq. (7), the frequency resolution is mainly determined by the sweep period $T$ of the chirped optical signal (V-shape LFM signal) and has no relation with the duty cycle, so it is studied without introducing the duty-cycle-enabling technique. The corresponding analysis bandwidth and sweep period $T$ are 63.2 GHz and 3.4375 ns. In the experiment, the two-tone SUT has a fixed frequency at 10 GHz, and the other frequency is set to 8.5 GHz, 8.7 GHz, 8.9 GHz, and 9 GHz. The corresponding time-frequency analysis results and waveforms are shown in Fig. 5. When the frequency interval of the two tones is 1 GHz, it is hard to distinguish the two frequencies in the time-frequency diagram. However, when the frequency interval is increased to 1.1 GHz, the two frequencies can be roughly distinguished. When the frequency interval is greater than 1.1 GHz, it is easy to distinguish the two frequencies. Therefore, a frequency resolution of 1.1 GHz can be achieved in the proposed system. According to Eq. (7), when $T=3.4375$ ns, the theoretical frequency resolution should be around 0.36 GHz. The difference between theoretical and experimental values is mainly due to the bandwidth limitations of the PD and the OSC. In the experiment, we further change the 3-dB bandwidth of the OSC to show the influence.

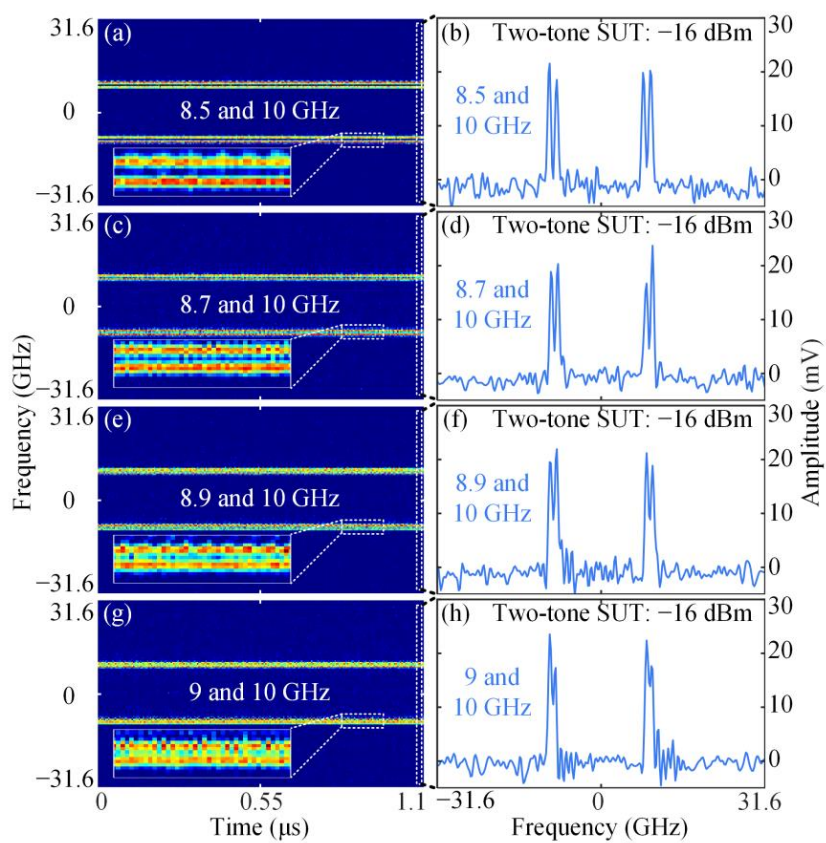

Fig. 5. Time-frequency analysis results and corresponding waveforms of the two-tone SUT when the two tones are set to (a), (b) 10 GHz and 8.5 GHz, (c), (d) 10 GHz and 8.7 GHz, (e), (f) 10 GHz and 8.9 GHz, and (g), (h) 10 GHz and 9 GHz.

Figure 6(a) shows the generated pulse waveforms of a single-tone signal at 15 GHz when the OSC 3-dB bandwidth is set to 20 GHz, 10 GHz, and 5 GHz, respectively, while Fig. 6(b) depicts the corresponding spectra via fast Fourier transform. As the OSC 3-dB bandwidth decreases, the FWHM of the generated pulse waveform increases, leading to a deterioration in the frequency resolution as shown in the corresponding time-frequency diagrams accumulated by 240 measurements in Fig. 6(c).

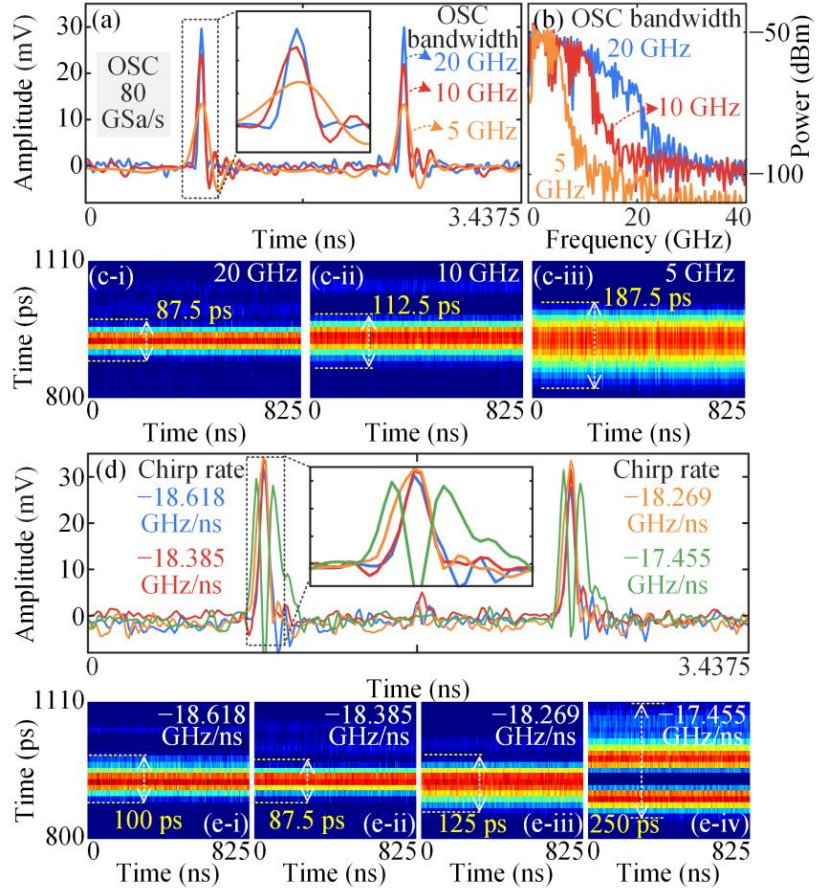

Fig. 6. (a) Pulse waveforms of a single-tone signal at 15 GHz under different OSC 3-dB bandwidths when the chirp rate of the V-shape LFM signal matches the dispersion of the DCF. (b) Spectra of the corresponding pulse waveforms in (a). (c) Time-frequency diagrams obtained through measurements over 240 sweep periods in (a). (d) Pulse waveforms of a single-tone signal at 15 GHz when the OSC 3-dB bandwidth is fixed at 20 GHz and the chirp rate of the V-shape LFM signal changes. (e) Time-frequency diagrams obtained through measurements over 240 sweep periods in (d).

As discussed in Section 2.2, the chirp rate of the V-shape LFM signal and the dispersion value of the DCF should be precisely matched to generate the compressed pulses. Here, the influence of the chirp rate of the V-shaped LFM signal on the frequency resolution is studied under the condition of a fixed dispersion value, that is, how the mismatch between the chirp rate and dispersion affects the frequency resolution. Figure 6(d) shows the generated pulse waveforms of a single-tone signal at 15 GHz when the OSC 3-dB bandwidth is set to 20 GHz and the chirp rate of the V-shape LFM signal is different, while Fig. 6(e) depicts the corresponding time-frequency diagrams accumulated by 240 measurements. It can be seen that the best FWHM of the pulses is obtained when the chirp rate is –18.385 GHz/ns, which is precisely matched with the dispersion. When the chirp rate changes, the FWHM increases, leading to a deterioration of the frequency resolution. When the chirp rate is –17.455 GHz/ns, there is even a situation where the pulse is split into two pulses. Therefore, it is essential to match the total dispersion of the DCF and the chirp rate of the V-shape LFM signal applied to the DP-MZM.

3.4 Temporal resolution and burst and interruption capturing

Temporal resolution is a key indicator reflecting the speed of time-frequency analysis, and it directly reflects the system's ability to capture burst signals. To further investigate the temporal resolution and the burst signal capturing capability of the proposed method, in this study, two parameters that influence the temporal resolution, i.e., the sweep period $T$ of the V-shape LFM signal and the duty cycle employed in the duty-cycle-enabling technique, are varied, while the chirp rate of the V-shape LFM signal is kept unchanged.

First, the temporal resolution is studied when the sweep period $T$ of the V-shape LFM signal is fixed and the duty cycle $p$ employed in the duty-cycle-enabling technique is changed. The SUT contains two components, one of which is a very short burst signal at 10 GHz, and the other is a single-tone signal at 15 GHz with a very short interruption, and the burst signal appears at the same time as the single-tone signal interrupts. The duration of the burst and interruption is set to 3.4375 ns, 6.875 ns, and 13.75 ns, respectively. The sweep period $T$ of the V-shape LFM signal is 3.4375 ns, while the duty cycle employed in the duty-cycle-enabling technique is set to 1, 1/2, and 1/4, respectively. Therefore, the corresponding theoretical temporal resolution is 3.4375 ns, 6.875 ns, and 13.75 ns, respectively. The corresponding time-frequency analysis results are shown in Fig. 7. As shown in Figs. 7(a)–(c), when the temporal resolution is 3.4375 ns, three bursts of different lengths are all well captured since they are all not less than 3.4375 ns. However, the interruption of 3.4375 ns is not fully captured, although the brightness of the time-frequency diagram at the interruption point decreases significantly. Furthermore, it is also evident that there is a duration difference between the bursts or interruptions captured in the time-frequency diagram and their actual lengths: The bursts shown in the diagram are longer by one sweep period than their actual values, while the interruptions are shorter by one sweep period. This is theoretically correct. Since it is challenging to align a burst signal perfectly with the sweep period in time in practice, a burst signal may span over an additional sweep period, causing it to be analyzed within that extra sweep period. Consequently, the length obtained in the time-frequency diagram is longer than its actual length. Similarly, if the interruption is not aligned with the sweep period, signals on the left and right sides of the interruption will be analyzed during the interruption, resulting in the actual obtained interruption length being shorter than its actual length. When the duty-cycle-enabling technique is introduced, if the lengths of bursts and interruptions are not shorter than the theoretical temporal resolution, Figs. 7(e), (f), and (i) can still be explained using the above theory.

It's important to note that in Figs. 7(d) and (h), the burst lengths are only half of the theoretical temporal resolution, yet the bursts are still detected. However, in Fig. 7(g), the burst is not detected. In fact, due to the implementation of the duty-cycle-enabling technique, if the duty cycle is $p$, only $p$ part of the temporal-domain signal is analyzed within the time duration corresponding to a single temporal resolution. Therefore, if a burst occurs within this analysis time window, it will be analyzed and displayed on the time-frequency diagram, as illustrated in Figs. 7(d) and (h). Conversely, if the burst does not fall within this analysis window, it will not be analyzed, as shown in Fig. 7(g). It is important to note that if the lengths of interruptions are shorter than the theoretical temporal resolution, as depicted in Figs. 7(d), (g), and (h), the interruptions cannot be captured.

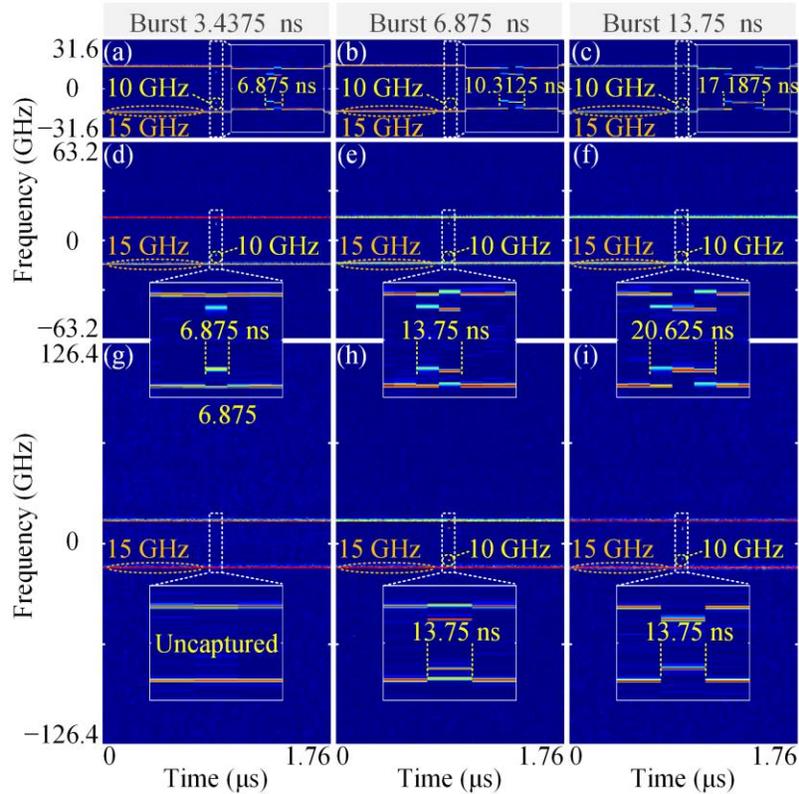

Fig. 7. Time-frequency analysis results of a burst signal at 10 GHz and an interrupted single-tone signal at 15 GHz. The duration of the burst and interruption is (a), (d), (g) 3.4375 ns, (b), (e), (h) 6.875 ns, and (c), (f), (i) 13.75 ns. The duty cycle employed in the duty-cycle-enabling technique is (a), (b), (c) 1, (d), (e), (f) 1/2, and (g), (h), (i) 1/4.

Figure 8 shows the time-frequency analysis results when the analysis bandwidth is decreased to 34.48 GHz and 17.24 GHz. The theoretical temporal resolution is 1.875 ns and 0.9375 ns. The SUT here is a very short burst signal at 2 GHz and a single-tone signal at 6 GHz with a very short interruption. The results obtained from Fig. 8 are also consistent with the conclusions drawn from our analysis of Fig. 7. Notably, in the results of Figs. 8(d)–(e), the time corresponding to the bursts and interruptions falls more precisely within the smallest integer multiple of the temporal resolution. Consequently, the time-frequency diagrams of the bursts and interruptions captured do not exhibit the situation of being one sweep period more or less, as observed in Fig. 7.

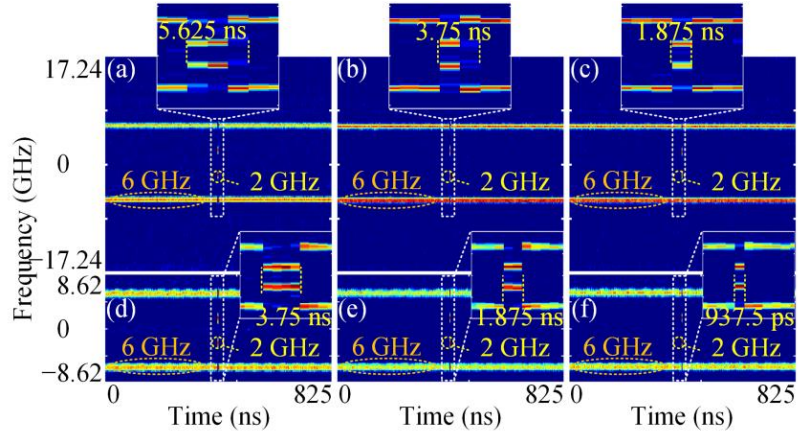

Fig. 8. Time-frequency analysis results of a burst signal at 2 GHz and an interrupted single-tone signal at 6 GHz. The duration of the burst and interruption is (a), (d) 3.75 ns, (b), (e) 1.875 ns, and (c), (f) 0.9375 ns. The temporal resolution is (a)–(c) 1.875 ns and (d)–(f) 0.9375 ns.

3.5 Power dynamic range

Finally, the power dynamic range of the proposed method is investigated. The system parameters are set to those used when obtaining Figs. 2 (c)–(d). The SUT is a single-tone signal generated by a microwave signal generator (MSG, Agilent 83630B). Figure 9 shows the measurement results at different SUT power when the frequency of the SUT is 5 GHz and 15 GHz. It can be seen that the minimum measurable power of the proposed system is around −27 dBm when the EA1 is employed, whereas the minimum measurable power of the proposed system is around 4 dBm when the EA1 is not employed. In the system shown in Fig. 1, only one stage of the amplifier is used. In practical applications, multiple stages of amplifiers can be added to the receiving end, along with the introduction of variable gain amplifiers, to further enhance the system's minimum measurable power and power dynamic range.

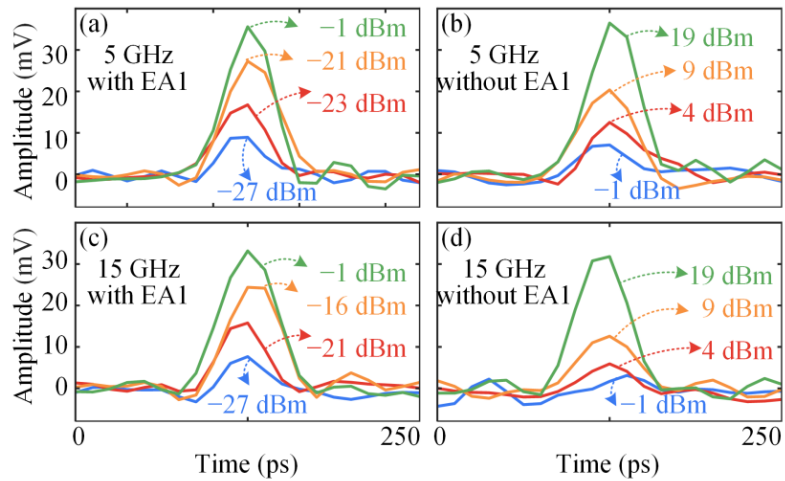

Fig. 9. Measurement results at different SUT power with and without employing EA1 when the frequency of the single-tone SUT is (a), (b) 5 GHz and (c), (d) 15 GHz.

## 4. Discussion

4.1 Enhancing temporal resolution and mitigating missing capture issues

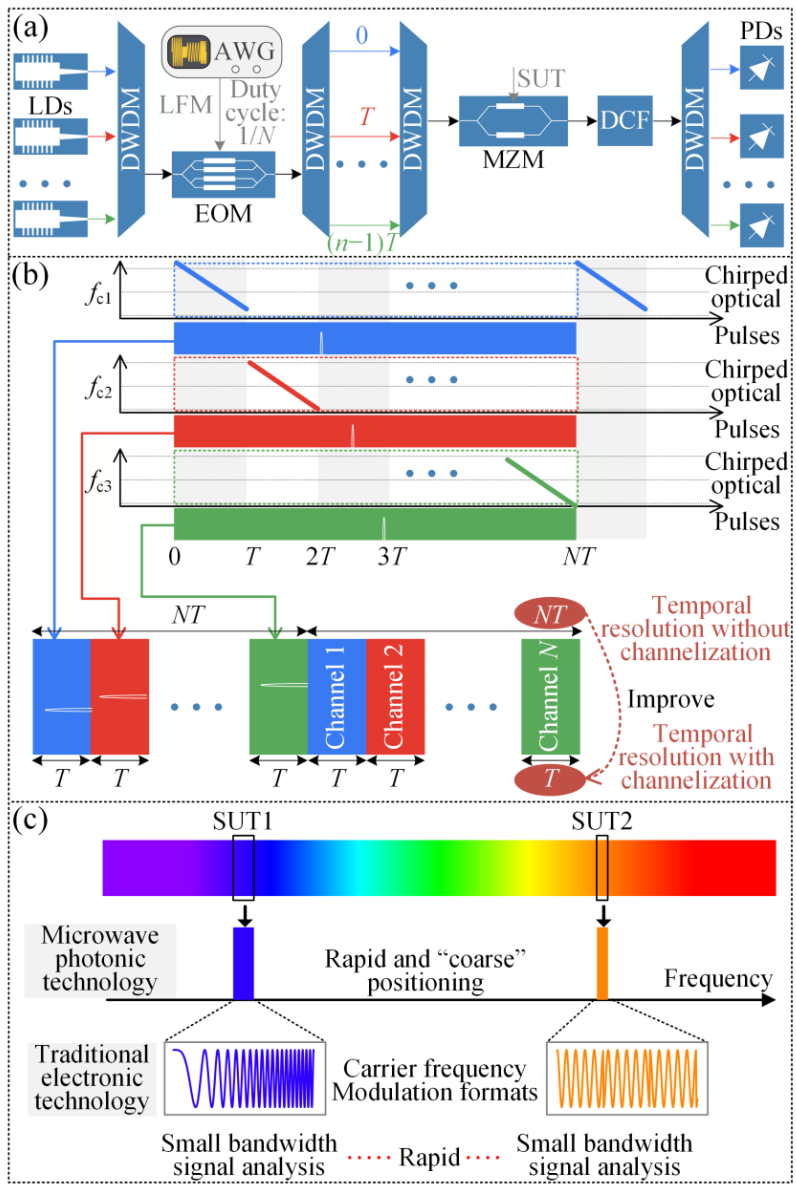

Fig. 10. (a) A temporal resolution improvement method based on wavelength division multiplexing when the duty-cycle-enabling technique is employed. (b) Principle of temporal resolution improvement in (a). (c) Principle of combining microwave photonic and traditional electronic-based methods for rapid, broadband, and accurate spectrum sensing. EOM, electro-optic modulator; DWDM, dense wavelength division multiplexing.

As investigated in Fig. 7, although the analysis bandwidth can be extended by applying the duty-cycle-enabling technique, there is temporal resolution deterioration and a possibility of missing capturing a burst or interruption whose duration is shorter than the temporal resolution. To address these issues, a channelization scheme can be employed, which allocates signals occurring at different times within an original temporal resolution $T/p$ to distinct channels for analysis, thereby enhancing the temporal resolution and mitigating the missing capture issues. For the duty cycle $p = 1/N$ of the V-shape LFM signal, a wavelength multiplexing scheme can be employed, as shown in Fig. 10 (a). $N$ Different wavelengths are combined by a dense wavelength division multiplexing (DWDM) multiplexer, and then all wavelengths are converted to chirp optical signals by the V-shape LFM signal with duty cycle 1/*N*. Then a pair of the DWDM multiplexer and demultiplexer, as well as a delay line matrix, is used to introduce different time delays to different channels. The delay of the delay matrix increases at equal intervals, which is the sweep period *T* of the V-shape LFM signal. When the duty cycle is 1/*N*, the temporal resolution of the single channel system shown in Fig. 1 is *NT*, while the temporal resolution of the channelized structure shown in Figs. 10(a) and (b) is still *T*. The cost of increasing temporal resolution is a more complex system structure.

4.2 Comparison with existing methods

A comparison between the proposed method and the previously reported microwave photonic time-frequency analysis methods is given in Table 1. The main advantages of the proposed method are as follows: (1) The ambiguity-free analysis bandwidth is only lower than that of [29], and the requirements for AWG and OSC are much lower than that of [29, 30] with similar bandwidths. In addition, without increasing the hardware requirements of the system, it is capable of achieving a bandwidth far exceeding 252.8 GHz demonstrated in this work and 448 GHz demonstrated in [29] through the duty-cycle-enabling technique; (2) It has very good reconfigurability on temporal resolution and analysis bandwidth by changing the duty cycle of the V-shape LFM signal; (3) Besides lower requirements for AWG and OSC, it also has a relatively lower requirement for modulator half-wave voltage and signal amplitude compared to the methods in [29] and [30]: the signal amplitude can be lower than $V_\pi$ in this work, while it should be $7V_\pi$ and $2V_\pi$ in [29] and [30], respectively.

Table 1. Comparison of different microwave photonic time-frequency analysis methods

| | Main hardware used for introducing a quadratic phase | Ambiguity-free analysis bandwidth | Frequency resolution | Temporal resolution | Reconfigurability | Absolute dispersion value @ wavelength | OSC bandwidth |
|---|---|---|---|---|---|---|---|
| Ref. [24] | MLL, DM | Unspecified | Unspecified | Unspecified | Low | 613.91 ps/nm @ 1551.22 nm | 63 GHz |
| Ref. [25] | CW LD, PM, RF signal source | 1.98 GHz | 60 MHz | 6.25 ns | Middle | 2000 ps/nm @ 1550 nm | 12 GHz |
| Ref. [26] | MLL | 2.43 GHz | 340 MHz | 5 ns | Low | 5351.06 ps/nm @ 1550 nm | 63 GHz |
| Ref. [28] | CW LD, AWG (50 GSa/s), PM | 0.53 GHz | 30 MHz | 30 ns | Middle | 450000 ps/nm @ - | 20 GHz |
| Ref. [29] | CW LD, PM, AWG (120 GSa/s) | 448 GHz | 16 GHz | 62.5 ps | High | 20.39 ps/nm @ 1550 nm | 500 GHz |
| Ref. [30] | CW LD, PM, AWG (92 GSa/s) | 92 GHz | 2.2 GHz 366 MHz | 1.5 ns 9.1 ns | High | 12152.59 ps/nm @ 1550 nm | 28 GHz |
| This work | CW LD, DP-MZM, AWG (64 GSa/s) | 252.8 GHz[*] 63.2 GHz 17.24 GHz | 1.1 GHz | 13.75 ns 3.3475 ns 0.9375 ns | High | 6817 ps/nm @ 1546.6 nm | 20 GHz |

[*] The analysis bandwidth can be further extended to THz, theoretically, by introducing a proper duty cycle on the V-shape LFM signal.

4.3 Consideration in practical applications

The microwave photonic frequency measurement and time-frequency analysis methods can overcome the bottleneck of traditional electronic-based methods, especially with significant advantages in real-time performance and analysis bandwidth. It has great application potential and space in applications with extremely high requirements for real-time performance and bandwidth. Compared with traditional electronic solutions, the main limitation of microwave photonic methods lies in the frequency measurement accuracy and frequency resolution, which are determined by the fundamental mechanisms and characteristics of optical and electronic systems. Therefore, a possible solution in the future is to combine microwave photonic technology with traditional electronic technology, as shown in Fig. 10(c). Utilizing the microwave photonic methods with high real-time performance and large analysis bandwidth, the frequency or time-frequency information of the SUT can be quickly and "roughly" located within a wide frequency band. Then, the frequency band where the signal exists can be down-converted to the intermediate frequency band by a rapidly agile local oscillator, and further high-accuracy and deep analysis can be performed using traditional electronic methods. Due to the "rough" frequency or time-frequency information provided by the microwave photonic methods, the electronic scheme does not need to blindly scan the entire wide frequency band, and the time consumed by local oscillator agility is greatly reduced. In addition, depending on the actual situation, the bandwidth of intermediate frequency analysis may also be significantly reduced based on the "rough" frequency or time-frequency information. Combining these two methods can not only achieve significantly better real-time analysis performance than traditional electronic solutions but also enable finer and deeper spectrum analysis than relying solely on microwave photonic methods.

In practical applications, we can further reduce the hardware requirement (for example the OSC bandwidth), by sacrificing a certain degree of temporal resolution. When the temporal resolution deteriorates, the chirp rate of the chirped signal needs to be decreased to ensure the same analysis bandwidth. Thus, according to Eq. (3), the dispersion value should be increased to compress the chirped optical signal. In the experiment, the dispersion is introduced by a long DCF, which has an insertion loss of around 30 dB, so two EDFAs are used before and after the DCF to amplify the optical signal after modulation and compensate for the loss of the DCF. Due to the large volume and high loss of the long DCF, in practical applications, linearly chirped fiber Bragg gratings [30] can be used as a substitute for DCF to significantly reduce the insertion loss while achieving the same large dispersion, thereby improving system performance.

5. Conclusion

In summary, we have introduced a microwave photonic frequency measurement and time-frequency analysis method with an ambiguity-free analysis bandwidth spanning hundreds of GHz, coupled with a GHz-level frequency resolution and a temporal resolution of 10 ns. The main contribution of this work lies in the introduction of a specially designed V-shape LFM signal and a duty-cycle-enabling technique, which enables the dispersion- and FTTM-based microwave photonic frequency measurement and time-frequency analysis method to achieve an ambiguity-free analysis bandwidth exceeding 200 GHz while utilizing hardware with lower requirements than existing microwave photonic solutions. Furthermore, it possesses the capability to scale up to THz bandwidth through the application of the duty-cycle-enabling technique without increasing hardware complexity. An experiment is performed. An ambiguity-free analysis bandwidth of 252.8 GHz is achieved by employing a duty cycle of 1/4 on the V-shape LFM signal, the corresponding temporal resolution is 13.75 ns, and the frequency resolution is 1.1 GHz. This research holds significant importance for real-time broadband spectrum sensing. It enables the acquisition of complex electromagnetic spectrum information within a bandwidth of hundreds of GHz, with an extremely high real-time capability down to the order of 10 nanoseconds. The proposed method can be applied to future cognitive radio, intelligent transportation systems, and especially electronic warfare systems that have stringent requirements for bandwidth and real-time performance.

**Acknowledgements**
National Natural Science Foundation of China (62371191, 62401207); Key Laboratory of Radar Imaging and Microwave Photonics (Nanjing University of Aeronautics and Astronautics), Ministry of Education (NJ20240004), Shanghai Aerospace Science and Technology Innovation Fund (SAST2022-074), Science and Technology Commission of Shanghai Municipality (22DZ2229004).